\documentclass[twocolumn,showpacs,preprintnumbers,amsmath,amssymb,10pt]{revtex4}
\usepackage{amsmath,epsfig,graphicx,amssymb}
\begin{document}

\title{STIRAP transport of Bose-Einstein condensate in triple-well trap}
\author{V.O. Nesterenko$^1$, A.N. Novikov$^{1}$, F.F. de Souza Cruz$^2$,
E.L. Lapolli$^2$}
\date{\today}
\address{$^{1}$
Bogoliubov Laboratory of Theoretical Physics, Joint Institute for Nuclear
Research, Dubna, Moscow region, 141980, Russia}
\affiliation{$^{3}$
Departamento de Fisica, Universidade Federal de Santa Catarina, Florianopolis,
SC, 88040-900, Brasil}

\begin{abstract}
The irreversible transport of multi-component Bose-Einstein condensate (BEC)
is investigated within the Stimulated Adiabatic Raman Passage
(STIRAP) scheme. A general formalism for a single BEC
in M-well trap is derived and analogy between multi-photon and tunneling
processes is demonstrated. STIRAP transport of BEC in a cyclic triple-well
trap is explored for various values of detuning and interaction between
BEC atoms. It is shown that STIRAP provides a complete population transfer
at zero detuning and interaction and persists at their modest values.
The detuning is found not to be obligatory. The possibility of non-adiabatic
transport with intuitive order of couplings is demonstrated. Evolution of the
condensate phases and generation of dynamical and geometric phases are inspected.
It is shown that STIRAP allows to generate the unconventional geometrical phase
which is now of a keen interest in quantum computing.
%STIRAP transport is compared with alternative non-adiabatic transfer.

\end{abstract}

\pacs{03.75.Mn, 03.75.Lm, 05.60.Gg}

\maketitle

\section{Introduction}

In recent decades, investigation of Bose-Einstein condensate (BEC) has
become one of the main streams in modern physics (see for reviews and
monographs \cite{Pit_Str_03,Petrick_Smith_02,Dalfovo_99,Leggett_01,
Yuk_01,Pitaevskii_UFN_06,crossover}).
In particular, a large attention is paid to dynamics of multi-component BEC
and related coherent phenomena, e.g the Josephson effect in weakly bound BECs
\cite{Yuk_JPA_96}-\cite{Opat_arXiv_08}.

Two kinds of multi-component BEC
%different by construction but similar by physics,
are usually considered. The first one is confined
in a single trap and  contains atoms in a few hyperfine levels
(multi-level system or MLS).
Here every component is formed by atoms in a given level.
The components can be coupled by the laser light with the
carrier frequency equal or close to the difference of the Bohr
frequencies of the hyperfine states. One can control the coupling
by varying parameters of the laser irradiation and so get
different regimes of the transfer of atoms between the components:
Josephson oscillations (JO), macroscopic quantum self-trapping (MQST),
etc (see e.g. \cite{Raghavan_PRA_99} for discussion).

In the second kind of the multi-component BEC, the atoms are
in the same hyperfine state but the trap is separated by
laser-produced barriers into a series of weakly bound wells
(multi-well system or MWS) \cite{arrays}. BEC atoms can tunnel through
the barriers and exhibit the similar effects as the MLS.
In this case BEC components are represented by populations of the wells.
Both MLS and MWS are obviously similar. Indeed,
the coupling Rabi frequencies in MLS are counterparts of
the barrier transition matrix elements in MWS. And detuning between Bohr
and carrier frequencies in MLS is similar to detuning of the well
depths in MWS. BEC in optical lattice \cite{Morsh_RMP_06,Yin_PR_06} can
be also treated as MWS though, unlike a few-well traps \cite{arrays},
the well depths and separations in optical lattices cannot be usually monitored
individually for every cell.

Most of the studies consider BEC with two components
\cite{Yuk_JPA_96}-\cite{Ne_lanl_08} and much less with three components
\cite{Ne_lanl_08}-\cite{Opat_arXiv_08}.
The later case is much more complicated. At the same time,
it promises  new dynamical regimes \cite{Nemoto_PRA_00,Zhang_PLA_01}
and allows to consider not only linear (couplings 1-2, 2-3)
but also cyclic (couplings 1-2, 2-3, 3-1) well chains.

In the present paper, we investigate BEC dynamics in the triple-well
trap, i.e. MWS with the number of wells M=3. Unlike most of the previous
studies, we will
explore not {\it oscillating} fluxes of BEC but its complete and
{\it irreversible} transport between the initial and target wells.
For this aim, the coupling between BEC fractions (=components) will
be monitored in time (unlike the constant coupling for the Josephson-like
oscillations). Once being realized, BEC transport could open interesting
perspectives in many areas, e.g. in exploration of coherent topological
modes \cite{topol_1,Yuk_04} and  diverse geometric phases
\cite{Fuentes_PRA_02}-\cite{Balakrishnan_EPJD_06}. The later is
especially important since geometric phases are considered as promising
information carriers in quantum computing
\cite{Nayak_RMP_08_gp,Zhu_PRL_03_gp,Feng_PRA_07_gp}.

Due to similarity between MLS and MWS, one may try to apply for
BEC transport numerous developments of modern quantum optics, in
particular, adiabatic two-photon population transfer methods
\cite{Vitanov}. Between them the
stimulated Raman adiabatic passage (STIRAP) \cite{Vitanov,Berg} seems to be the
most suitable for our aims since it allows, at least in principle, the
{\it complete} irreversible population transfer. The method was first
developed for atoms and simple molecules \cite{Vitanov,Berg} and
then probed in metal clusters \cite{Ne_PRA_04}-\cite{Ne_book_06} and
variety of other systems, see references in \cite{Rab_PRA_08}.
Quite recently STIRAP was applied to the transport of individual atoms
\cite{Eckert_PRA_04} and BEC
\cite{Graefe_PRA_06,Ne_lanl_08,Ne_Bars_08,Rab_PRA_08,Opat_arXiv_08}
in the triple-well trap.

The applicability of STIRAP to BEC transport needs some care since
interaction between BEC atoms results in a time-dependent nonlinearity
of the problem,  which can destroy the adiabatic transfer
\cite{Graefe_PRA_06,Ne_Bars_08,Rab_PRA_08,Opat_arXiv_08}.
This nonlinearity plays the same detrimental role as the
dynamical Stark shift in electronic MLS systems, where it
disturbs the two-photon resonance condition and thus breaks
one of the basic STIRAP requirements (see discussion in Sec.
\ref{sec:STIRAP}). As was shown in \cite{Graefe_PRA_06},
the undesirable nonlinear impact can be circumvented by using a detuning
larger than the atomic interaction. The subsequent studies
\cite{Rab_PRA_08,Opat_arXiv_08} confirmed that the detuning is
useful if we  aim  a minimal (say $ <1\%$) occupation of
the intermediate well during STIRAP process. The less the (temporary)
occupation, the better adiabaticity and robustness of the process.
At the same time, one should recognize that occupation of the intermediate
well cannot be fully avoided. Moreover, that occupation is temporal and
in any case is further transferred to the final well. So it does not affect
the final fidelity of the BEC transport.

In this study we will show that the robust and
complete transport of the interacting BEC can take place even at zero detuning,
regardless of the temporary weak population of the intermediate state. Of course
such transport is more likely quasiadiabatic but we are interested in the
transport completeness rather than in its purely adiabatic character. Moreover,
we will show that the complete transfer can be done even at intuitive sequence
of the pump and Stokes couplings (unlike their counterintuitive order in STIRAP),
i.e. in strictly non-adiabatic case.

As compared with the previous studies \cite{Graefe_PRA_06,Rab_PRA_08,Opat_arXiv_08},
we will consider more general triple-well trap which has also 3-1 coupling and thus
represents the circular configuration \cite{Ne_lanl_08,Ne_Bars_08}. Such configuration
allows to run BEC through the circle as many rounds as we want
and put it to any of three wells. Besides the populations, the condensate
phases will be explored. Moreover, we will present some first examples of
the dynamical and geometric phases generated in STIRAP. The later is
possible because the circular well configuration and adiabatic STIRAP transfer
allow to build the adiabatic cyclic evolution. It worth noting that condensate
phases and their dynamical and geometric constituents were not yet explored
in STIRAP (for exception of a brief phase analysis in \cite{Ne_Bars_08}).

The paper is outlined as follows. In Sec. II we sketch STIRAP.
In Sec. III a general mean-field formalism for dynamics of
multi-component BEC is presented and specified for MWS with M=3.
In Sec. IV the calculation scheme is given and similarity between
our scenario and conventional STIRAP is discussed. In Sec. V
results of the calculations are discussed.
The conclusions are done in Sec. VI.

\section{STIRAP}
\label{sec:STIRAP}

\begin{figure}
\includegraphics[height=8cm,width=2.4cm,angle=-90]{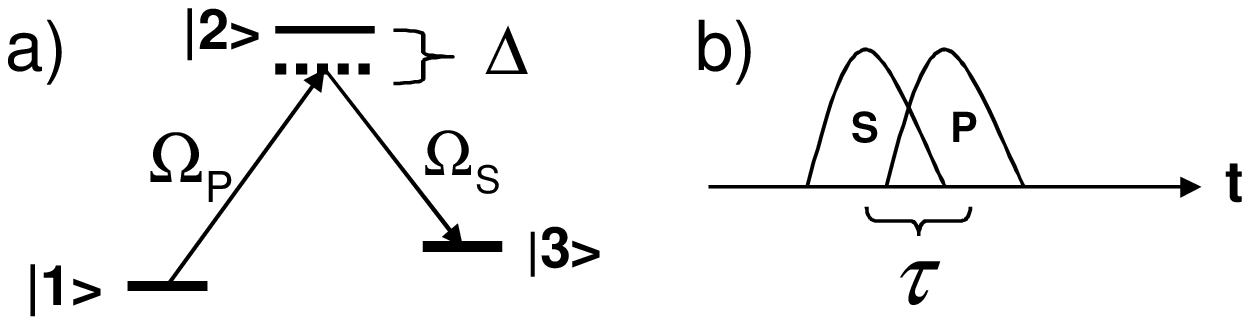}
\caption{\label{fig1_STIRAP}
a) STIRAP scheme for a three-level $\Lambda$-system.
The pump and Stokes pulses with Rabi frequencies $\Omega_P(t)$
and $\Omega_S(t)$ provide the couplings $|1\rangle-|2\rangle$ and
$|2\rangle-|3\rangle$,
respectively. $\Delta$ is the detuning from the intermediate level.
b) Counterintuitive sequence of pump and Stokes pulses overlapping
for a time $\tau$.
}
\end{figure}

STIRAP \cite{Vitanov,Berg} is the adiabatic two-step process providing the complete
population transfer from the initial level $|1\rangle$ to the target level
$|3\rangle$ via the intermediate level $|2\rangle$.
Its scheme for MLS is illustrated in Fig. \ref{fig1_STIRAP}.
%%%%%%%%%%%%
% Figure 1
%%%%%%%%%%%%
As is seen, the transfer is driven by the
pump and Stokes laser pulses with Rabi frequencies $\Omega_P(t)$
and $\Omega_S(t)$. The pump laser couples the initial and
intermediate states while the Stokes laser stimulates
the emission from the intermediate state to the target one.
In addition to the $\Lambda$-configuration given in the plot a),
STIRAP can be also realized in the ladder
and $\nu$-configurations \cite{Vitanov}.

STIRAP has three principle requirements:\\
1) two-resonance condition for the laser carrier and Bohr
frequencies
\begin{equation}\label{eq:two-resonance_cond}
\omega_P-\omega_S=\omega_3-\omega_1
\end{equation}
allowing a detuning
$\Delta=(\omega_2-\omega_1)-\omega_P=(\omega_2-\omega_3)-\omega_S$
from $\omega_2$;\\
2) overlap of the pulses (during the time $\tau$) and
their counterintuitive order (the Stokes pulse
proceeds the pump one);\\
3) adiabatic evolution ensured  by the condition
\begin{equation}\label{eq:adiab_cond}
\Omega \tau > 10
\end{equation}
where $\Omega=\sqrt{\Omega_P^2+\Omega_S^2}$.

Due to interaction with the laser irradiation,
the bare states $|1\rangle$, $|2\rangle$, $|3\rangle$ are
transformed to the dressed states
\begin{eqnarray}
\nonumber
|a^+\rangle&=&\sin\bar\theta \sin\bar\phi |1\rangle
+\cos\bar\phi |2\rangle
+\cos\bar\theta \sin\bar\phi |3\rangle ,\\
\label{eq:dressed_states}
|a^0\rangle&=&\cos\bar\theta |1\rangle-\sin\bar\theta |3\rangle ,\\
\nonumber
|a^-\rangle&=&\sin\bar\theta \cos\bar\phi |1\rangle
-\sin\bar\phi |2\rangle
+\cos\bar\theta \cos\bar\phi |3\rangle
\end{eqnarray}
with the spectra
\begin{equation}
  \omega^{\pm}=\Delta\pm\sqrt{\Delta^2+\Omega^2},
  \quad
  \omega^0=0
\end{equation}
and mixing angle $\bar\theta$ determined through
\begin{equation}\label{eq:sin_cos_theta}
  \sin\bar\theta =\Omega_P/\Omega ,\quad \cos\bar\theta =\Omega_S/\Omega .
\end{equation}
The mixing angle $\bar\phi$, a known function of the Rabi frequencies and
detuning \cite{Gaubatz_JCP_90}, is of no relevance in the
present discussion. For the sake of simplicity, we omitted
in this section time dependence of Rabi frequencies and other values.

STIRAP has been developed for population of non-dipole states (with the
spin $J \ne 1$) which cannot be excited in the photoabsorption
but can be reached by two dipole transitions
via an intermediate state. The main aim was to avoid, as much as
possible, the population of the intermediate state and thus prevent
the leaking via decay of this state. As it  is seen from Eq.
(\ref{eq:dressed_states}), this can be achieved by keeping the
system during all the process in the dressed state $|a^0\rangle$
which has no contribution from the intermediate
state $|2\rangle$. For this aim we need an adiabatic evolution following
with the counterintuitive sequence of the pump
and Stokes pulses as is shown in Fig. 1b. In this case, we have
$\Omega_S(t) \ne 0, \Omega_P(t)=0, \sin\bar\theta =0, \cos\bar\theta =1$
at early time and
$\Omega_S(t)=0, \Omega_P(t) \ne 0, \sin\bar\theta =1, \cos\bar\theta =0$
at late time. Then,  unlike $|a^+\rangle$ and $|a^-\rangle$,
the state $|a^0\rangle$ is reduced to $|1\rangle$ at the
beginning and to $|3\rangle$ at the end of the evolution.
The main point is to evolve
the system adiabatically, keeping it all the time in the state
$|a_0\rangle$. As is seen from (\ref{eq:adiab_cond}), for this aim
we need either a strong coupling $\Omega$ (i.e. high laser intensities)
or a long overlapping time $\tau$ which is just  duration of the
adiabatic evolution.

The STIRAP Hamiltonian reads
\begin{equation}\label{eq:H_STIRAP}
\bar{H}(t)=\frac{\hbar}{2}
\left(\begin{array}{l} {\;\; 0 \qquad \Omega_P \qquad 0} \\
{\Omega_P \quad 2\Delta \quad \Omega_S }\\
{\;\; 0 \qquad \Omega_S \qquad 0}
\end{array} \right) .
\end{equation}

Note that eqs. (\ref{eq:dressed_states})-(\ref{eq:H_STIRAP})
are obtained in the rotating wave approximation
(RWA). Using the fact that
$\omega_{P,S},\omega_{1,2,3} \gg \Omega_{P,S}, \Delta$,
the RWA allows to omit the high laser and Bohr frequencies
and keep only the low frequencies of interest.

Eqs. (\ref{eq:dressed_states})-(\ref{eq:H_STIRAP})
also neglect the dynamical Stark shifts $\delta_i(t)$ pertinent to MLS,
which are supposed here to be weak, $\delta_i(t) \ll \Omega_{P,S}, \Delta$.
Otherwise the shifts enter the diagonal terms in (\ref{eq:H_STIRAP})
and complicate Eqs. (\ref{eq:dressed_states})-(\ref{eq:sin_cos_theta})
\cite{Vitanov}.
The dynamical Stark shifts are detrimental for STIRAP since they
destroy the two-resonance condition (\ref{eq:two-resonance_cond}).
The larger Rabi frequencies $\Omega_{P,S}$ (and hence the laser intensities),
the stronger the Stark shifts. So, $\Omega_{P,S}$ must be large enough
to keep the adiabatic condition (\ref{eq:adiab_cond}) and, at
the same time, small enough not to cause too strong Stark shifts.
This problem is obviously absent in MWS.  As was mentioned above, the
time-dependent nonlinearity in BEC is detrimental for STIRAP
\cite{{Graefe_PRA_06}}. In fact it plays the similar
destructive role as the Stark shift.

\section{Mean-field description of BEC dynamics in MWS}

We start from the non-linear Schr$\ddot o$dinger, or Gross-Pitaevskii
equation \cite{GPE}
\begin{equation}
\label{eq:NLSE}
i\hbar{\dot \Psi}({\vec r},t) = [-\frac{\hbar^2}{2m}\nabla^2
+ V_{ext}({\vec r},t)
+ g_0|\Psi({\vec r},t)|^2]\Psi({\vec r},t)
\end{equation}
where the dot means time derivative, $\Psi({\vec r},t)$ is the classical
order parameter of the system, $V_{ext}({\vec r},t)$ is the external trap
potential involving both (generally time-dependent) confinement and coupling,
$g_0=4\pi a/m$ is the parameter of interaction
between BEC atoms, $a$ is the scattering length and $m$ is the atomic
mass.

In what follows, we will consider the MWS where BEC is distributed
between M wells separated by barriers. Then BEC components are reduced
to the condensate fractions (=populations) in the wells.
For BEC with weakly bound M fractions, the order parameter can be expanded as
\cite{Raghavan_PRA_99}
\begin{equation}\label{eq:Psi}
\Psi({\vec r},t)=\sqrt{N}\sum_{k=1}^M \psi_k(t)\Phi_k({\vec r})
\end{equation}
where $\Phi_k({\vec r})$ is the static ground state solution of
(\ref{eq:NLSE}) for the isolated (without coupling) k-th well
\cite{note} and
\begin{equation} \label{order_par}
\psi_k(t)=\sqrt{N_k(t)}e^{i\phi_k(t)}
\end{equation}
is the amplitude related with the relative population $N_k(t)$
and corresponding phase $\phi_k(t)$ of the k-th fraction.
The total number of atoms $N$ is conserved:
$\int d\vec r |\Psi({\vec r},t)|^2/N=\sum_{k=1}^M N_k(t) = 1 \; .$

Being mainly interested in evolution of populations $N_k(t)$ and
phases $\phi_k(t)$,
we dispose by integration of the spatial distributions
$\Phi_k({\vec r})$  and finally get
\cite{Smerzi_PRL_97,Raghavan_PRA_99}
\begin{equation}\label{psi(t)}
  i{\dot \psi}_k = [E_k(t)+ UN|\psi_k|^2]\psi_k
- \sum_{j \ne k}^M \Omega_{kj}(t) \psi_j \;
\end{equation}
where
\begin{equation}\label{Om}
 \Omega_{kj} (t) = - \frac{1}{\hbar}\int d{\vec r} \;
 [\frac{\hbar^2}{2m}\nabla\Phi^*_k \cdot\nabla\Phi_j
 +\Phi^*_k V_{ext}(t)\Phi_j]
\end{equation}
is the coupling between BEC fractions,
\begin{equation}\label{E}
  E_k(t)= \frac{1}{\hbar}\int d{\vec r} \;
  [\frac{\hbar^2}{2m}|\nabla\Phi^*_k|^2
  +\Phi^*_k V_{ext}(t)\Phi_k]
\end{equation}
is the potential depth, and
\begin{equation}\label{U}
  U= \frac{g_0}{\hbar}\int d{\vec r} \; |\Phi_k|^4 \;
\end{equation}
labels the interaction between BEC atoms. The values $\Omega_{kj} (t)$,
$E_k(t)$, and $U$ have dimension of the frequency.

For simplicity we suppose that all the couplings have a common peak
amplitude $K$. Then it is convenient to pick out this  amplitude
from the couplings
\begin{equation}\label{Omega}
\Omega_{kj}(t)=K {\bar\Omega}_{kj}(t)
\end{equation}
and scale (\ref{psi(t)}) by $1/2K$ so as to get
\begin{equation}\label{spsi(t)}
  i\dot{\psi}_k = [\bar{E}_k(t)
  +\Lambda |\psi_k|^2]\psi_k
- \frac{1}{2}\sum_{j \ne k}^M \bar{\Omega}_{kj}(t) \psi_j \; .
\end{equation}
Here
\begin{equation}\label{Lambda}
\bar{E}_k(t)=E_k(t)/2K, \quad \Lambda=UN/2K
\end{equation}
and the time is scaled as $2Kt \to t$ thus becoming dimensionless.
Eq. (\ref{spsi(t)}) is convenient since it is driven by one key
parameter $\Lambda$ responsible for the interplay between the
coupling and interaction. As is shown below, this parameter is
decisive for STIRAP transport of BEC.

By substituting (\ref{order_par}) into (\ref{spsi(t)}) and separating
real and imaginary part, one gets equations describing evolution of the system
in terms of fractional  populations $N_k(t)$ and
phases $\phi_k(t)$:
\begin{eqnarray}\label{eq:N_dot}
  {\dot N}_k&=& -\sum_{j \ne k}^M \bar{\Omega}_{kj}\sqrt{N_j N_k}
 \sin (\phi_j-\phi_k) \; ,
\\
  {\dot \phi}_k &=& -[\bar{E}_k +\Lambda N_k]
+ \frac{1}{2}\sum_{j \ne k}^M \bar{\Omega}_{kj}\sqrt{\frac{N_j}{N_k}}
 \cos (\phi_j-\phi_k).
\label{eq:phi_dot}
\end{eqnarray}

In MWS, the condensate is distributed between M space-shifted coupled
wells, $V_{trap}(\vec r) \to \sum_{k=1}^M V_k(\vec r)$, with the depths
$E_k$. The wells are separated by the
barriers with penetrabilities $\Omega_{k\ne j}(t)$. We consider below
a weak coupling between BEC fractions. Then only the neighbor
fractions are coupled, $\Omega_{kj}\ne 0$ for $1 \le j=k\pm1 \le M$,
and interaction between atoms of different fractions can be neglected.

Equations (\ref{eq:N_dot})-(\ref{eq:phi_dot}) allow the classical analogy
with the populations $N_k(t)$ and phases $\phi_k(t)$
treated as classical canonical conjugates. It is easy to verify
that these equations can be casted in the canonical form
\begin{eqnarray}\label{eq:cl_N_dot}
 {\dot N}_k &=& - \frac{\partial {\rm H}_{cl}}{\partial \phi_k} \; ,
 \\
\label{eq:cl_phi_dot}
{\dot \phi}_k &=& \frac{\partial {\rm H}_{cl}}{\partial N_k}
\end{eqnarray}
with the classical Hamiltonian
\begin{eqnarray}\label{eq:class_ham}
 {\rm H}_{cl}= &-&\sum_k^M (\bar{E}_k N_k +\frac{\Lambda N_k^2}{2})
\nonumber\\
 &+&
 \frac{1}{2}\sum_{kj}^M \bar{\Omega}_{kj}\sqrt{N_k N_j} \cos (\phi_j - \phi_k) \; .
\end{eqnarray}

One may further upgrade (\ref{eq:N_dot})-(\ref{eq:phi_dot})
by means of  canonical transformation of $N_k$ and $\phi_k$ to canonical
pairs related to the population imbalances and phase differences.
This will allow to remove
from (\ref{eq:N_dot})-(\ref{eq:phi_dot}) the integral of motion $N$
and decrease the total number of equations from 2M to 2(M-1).

Let us consider the linear canonical transformation
\begin{equation}\label{eq:linear_canon_tr}
z_k=\sum_j^M T_{kj} N_j, \quad \theta_k=\sum_j^M R_{kj} \phi_j
\end{equation}
with $z_k$ and $\theta_k$ being the population imbalances and
phase differences, respectively.  For the linear transformation,
the matrices $T$ and $R$ in (\ref{eq:linear_canon_tr})
are related as
\begin{equation}
R={\tilde T}^{-1}
\end{equation}
i.e. the transformation matrix for the phases is the
inverse transposed matrix for the populations.

For the MWS case with M=3 it is natural to chose
\begin{equation}\label{eq:z_canon_tr}
z_1=N_2-N_1, \quad z_2=N_3-N_2, \quad z_3=N \; .
\end{equation}
Then one gets
\begin{equation}
\label{eq:T_matrix}
T=
\left(\begin{array}{ccc}
   -1   &\;1&  0 \\
 \; 0   & -1&  1 \\
 \; 1   &\;1&  1 \\
\end{array}\right),
%\end{equation}
%and
%\begin{equation}
%\label{eq:R_matrix}
\;
R= \frac{1}{3}
\left(\begin{array}{ccc}
  -2  &\; 1& 1 \\
  -1  &  -1& 2 \\
\; 1  &\; 1& 1 \\
\end{array}\right) .
\end{equation}
Matrices for the inverse transformation
\begin{equation}\label{eq:N_z}
N_k=\sum_j^M T^{-1}_{kj} z_j, \quad
\phi_k=\sum_j^M R^{-1}_{kj} \theta_j
\end{equation}
for M=3 read
\begin{equation}
\label{eq:R_matrix}
T^{-1}= \frac{1}{3}
\left(\begin{array}{ccc}
 -2  & -1&  1 \\
\;1  & -1&  1 \\
\;1  &\;2&  1 \\
\end{array}\right),
\;
R^{-1}=
\left(\begin{array}{ccc}
-1  &\;0&  1 \\
\;1  &-1&  1 \\
\;0  &\;1&  1 \\
\end{array}\right).
\end{equation}

New variables include the integral of motion
$z_3=N$ and the total phase $\theta_3=\phi_1+\phi_2+\phi_3$.
The equations for these values are straightforwardly  separated
from (\ref{eq:N_dot})-(\ref{eq:phi_dot}) and can be skipped.
Then it is enough to solve the remaining four equations
for $z_1$, $z_2$, $\theta_1$  and $\theta_2$.

Note that the formalism presented above is
general and can be applied to both i) {\it oscillating} BEC fluxes
in traps with {\it constant} parameters (like in
\cite{Yuk_JPA_96}-\cite{Nista_08} and
\cite{Nemoto_PRA_00}-\cite{Mossmann_PRA_06})
and ii) irreversible BEC transport in traps with time-dependent parameters,
e.g. $\Omega_{kj}(t)$. This formalism is partly given elsewhere
\cite{Yuk_JPA_96}-\cite{Opat_arXiv_08}. However, we find
useful to present here its full and consistent version.

%%%%%%%%%%%%
% Figure 2
%%%%%%%%%%%%
\begin{figure}
\includegraphics[height=8cm,width=3.0cm,angle=-90]{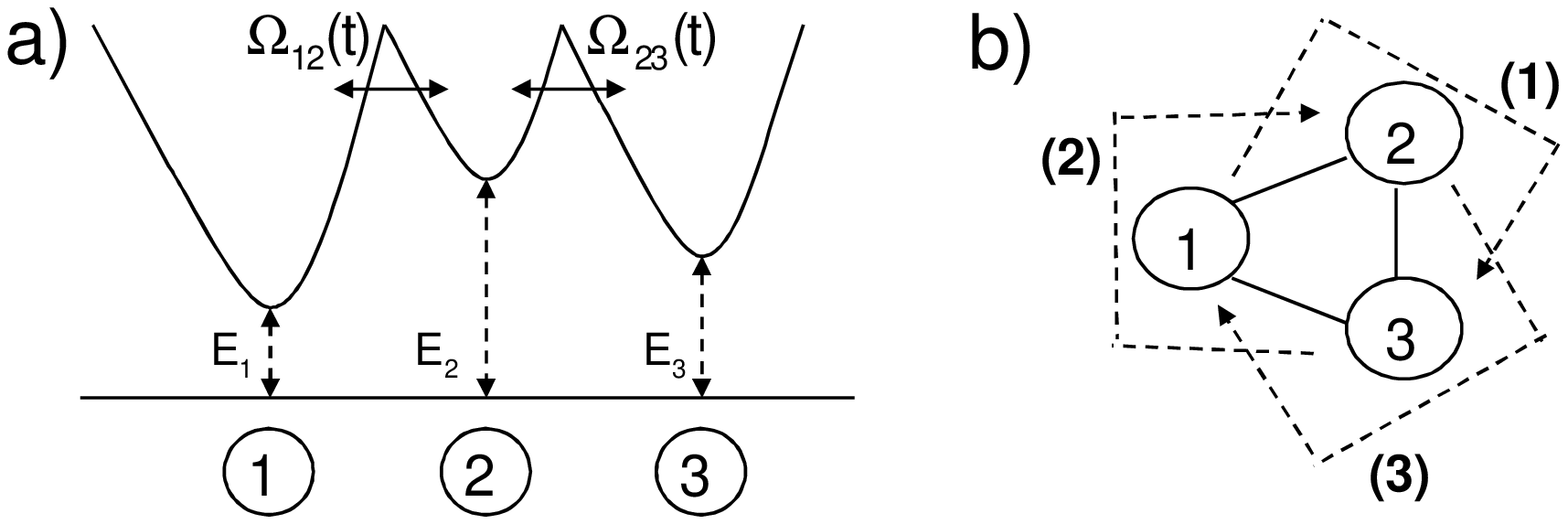}
\vspace{3mm}
\caption{\label{fig:stirap_tunnel}
a) The linear triple-well system containing the interacting BEC.
The wells are denoted by the numbered circles. The tunneling couplings
$\Omega_{12}(t)$ and $\Omega_{23}(t)$ are marked by double arrows.
$E_1$, $E_2$ and $E_3$ are well depths. b) The circular configuration
of triple-well system with the additional coupling 1-3. The solid lines
connecting the circles mean the couplings. The dash curves represent
the successive STIRAP transfers (1), (2) and (3), after which BEC
returns back to the initial well $|1\rangle$. At every transfer only two
couplings are active, as is shown in the plot a).
}
\end{figure}

\section{Calculation scheme for M=3}

The triple-well system with interacting BEC and scheme of the cyclic STIRAP
transport  used in this study are depicted in Fig. 2. As is seen from the plot a),
the adjoining wells are (weakly) coupled via the tunneling and the couplings
$\Omega_{12}$ and $\Omega_{23}$ play the role of the pump $\Omega_{P}$ and
Stokes $\Omega_{S}$ Rabi frequencies in a familiar STIRAP scheme in Fig 1a).
Further, the plot b) shows that, compared with other STIRAP applications
for the transport of individual atoms \cite{Eckert_PRA_04} and BEC
\cite{Graefe_PRA_06,Rab_PRA_08,Opat_arXiv_08},
we implement here three STIRAP transfers, $|1\rangle \to |3\rangle$,
$|3\rangle \to |2\rangle$ and $|2\rangle \to |1\rangle$,
and every transfer follows the scheme a). In other words,
these three transfers run via the intermediate states $|2\rangle$, $|1\rangle$, and
$|3\rangle$ and use pairs of the couplings $(\Omega_{12}, \Omega_{23})$,
$(\Omega_{31}, \Omega_{12})$ and $(\Omega_{23}, \Omega_{31})$, respectively.
As was mentioned above, we need three STIRAP steps to produce the cyclic
evolution of the system and generate a geometric phase. Besides, this allows
to test fidelity of STIRAP transport in the chain of transfers.

The similarity between the present scenario and typical STIRAP
can be additionally justified by the comparison of their Hamiltonians.
Following Fig. 2a), Eqs. (\ref{spsi(t)}) for M=3 can be written as
\begin{equation}
i\hbar {\dot{\psi}}_k (t) = H(t) {\psi}_k(t)
\end{equation}
with the Hamiltonian
\begin{equation}\label{eq:H_BEC}
H(t)=
%\frac{\hbar}{2}
\left(\begin{array}{l} {\bar{E}_1+A_{1}(t) \quad \;\;\Omega_P(t) \quad\quad\quad\quad 0} \\
{\quad\Omega_P(t) \quad\quad \bar{E}_2+A_{2}(t) \;\; \quad\Omega_S(t)}\\
{\quad 0 \quad\qquad\quad\quad  \Omega_S(t) \quad\quad \bar{E}_3+A_{3}(t)}
\end{array} \right) \;
\end{equation}
where $A_{k}(t)=\Lambda N_k(t)$ is the non-linear interaction contribution and
$\Omega_P(t)=-\bar{\Omega}_{12}(t)/2$ and $\Omega_S(t)=-\bar{\Omega}_{23}(t)/2$
are the pump and Stokes couplings. The nonlinear terms $A_{k}(t)$ are
detrimental for adiabatic transfer within STIRAP \cite{Graefe_PRA_06}.
If to omit them, then (\ref{eq:H_BEC}) fully coincides with STIRAP
Hamiltonian (\ref{eq:H_STIRAP}).

In our study the time-dependent part of the coupling (\ref{Omega})
has the Gauss form
\begin{equation}\label{Omega_temporal}
{\bar\Omega}_{kj}(t)=\exp\{-\frac{(t_{kj}-t)^2}{\Gamma^2}\}
\end{equation}
where $t_{kj}$ and $\Gamma_{kj}$ are centroid and
width parameters. This form is smooth which is important for
adiabaticity  of the process.

Using (\ref{Omega}) and (\ref{Omega_temporal}) one may amend
the STIRAP adiabatic condition (\ref{eq:adiab_cond}).
Following \cite{Vitanov,Messia} one gets
\begin{equation}\label{eq:adiab_BEC}
\frac{\Omega_S\dot{\Omega}_P-\Omega_P\dot{\Omega}_S}
{\Omega^3}\ll 1
\; \mbox{or} \; \Omega_{\tau} \gg \frac{|d|}{\sqrt{2}\; \Gamma^2}
\end{equation}
where $\Omega_{\tau}$ is the average amplitude of the pump and Stokes couplings
during the overlap time $\tau$ and  $d=t_{12}-t_{23}$ is the relative pump-Stokes
shift. Usually one may take $\Omega_{\tau} \approx 0.5K$ and
$\tau \approx \Gamma$. Then (\ref{eq:adiab_BEC}) is cast to
\begin{equation}\label{eq:adiab_BEC_2}
 K\tau^2 \gg \sqrt{2} |d|
\end{equation}
which means that STIRAP needs a strong coupling amplitude $K$
and/or a long overlap time $\tau$. It is easy to see that
(\ref{eq:adiab_BEC_2}) remains to be the same for the scaled
dimensionless time.

Since we use the mean-field approximation, the number of atoms in BEC
should be sufficiently large to neglect the quantum
corrections. In the present study we suppose that $N > 10^4$
\cite{Smerzi_PRL_97,Raghavan_PRA_99}.
The total number of atoms is included to the parameter
$\Lambda$ in (\ref{Lambda}).

\section{Results and discussion}
\subsubsection{Populations and phases}

Results of the calculations are depicted in Figs. 3-6. In all
the figures time is dimensionless. The coupling parameters are
$\Gamma$ = 5.4 and $d$ = -5. Then for $K=1$ we get
$K\Gamma^2 \approx$ 30 and $\sqrt{2} |d| \approx$ 7
and so keep the adiabatic condition (\ref{eq:adiab_BEC_2}).

In Fig. 3 the populations $N_i(t)$ during three STIRAP steps
are exhibited. The calculations are performed at initial
conditions (t=0) $N_1=1, \; N_2=N_3=0$ and $\phi_1=\phi_2=\phi_3=0$.
The sequence of the pairs of Stokes (first) and pump (second)
couplings is given in the panel a) while other plots demonstrate
evolution of $N_i(t)$ at different values of the ratio $\Lambda$
and detuning $\Delta$.

As is seen from the panel b), all three STIRAP steps are robust and complete
for $\Lambda=\Delta=0$, i.e. without interaction $U$ and detuning.
The minor peaks observed at t=20, 80 and 140, i.e. during the transfers,
mean a weak temporary population of the intermediate wells, which hints
that the transport is not fully adiabatic. However, what is important for
our aims, the transport is complete. And this takes place even after
3 STIRAP steps and despite deviations (though minor)  from adiabaticity.

%Figure 3
\begin{figure*}
\includegraphics[height=12cm,width=12.0cm,angle=-90]{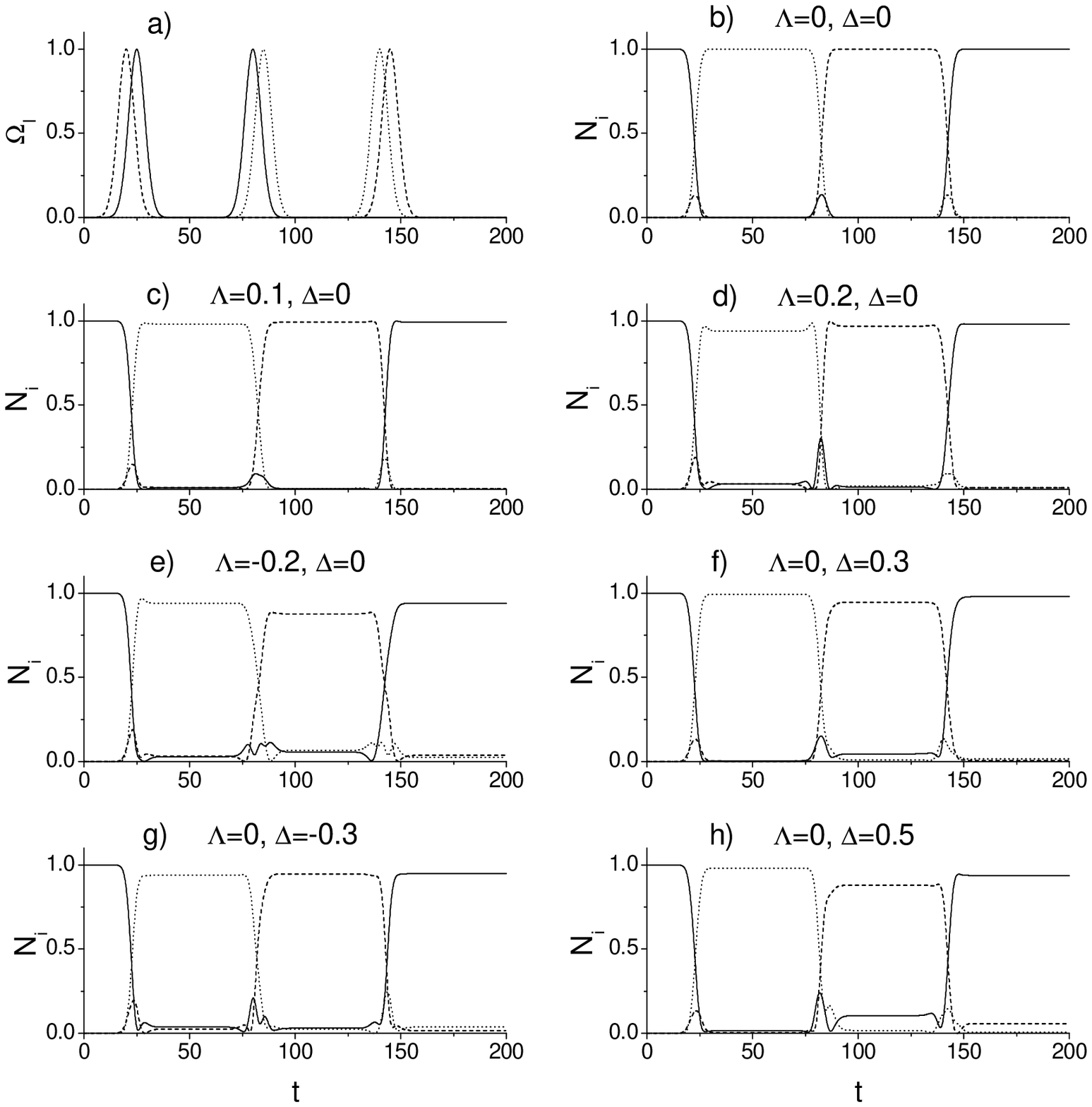}
\vspace{3mm}
\caption{\label{fig:populations}
STIRAP transport of BEC in the circular well chain.
a): Sequence of the pairs of Stokes (first) and pump (second)
couplings $\Omega_{kj}$ responsible for 1-2 (solid line), 2-3
(dash line) and 2-3 (dotted line) couplings. b)-h): Evolution
of the populations $N_1$ (solid line) $N_2$ (dash line), and $N_3$
(dotted line) at different values of the ratio $\Lambda$
and detuning $\Delta$.
}
\end{figure*}
%%%%%%%%%%%%
% Figure 4
%%%%%%%%%%%%
\begin{figure}
\includegraphics[height=8cm,width=4.5cm,angle=-90]{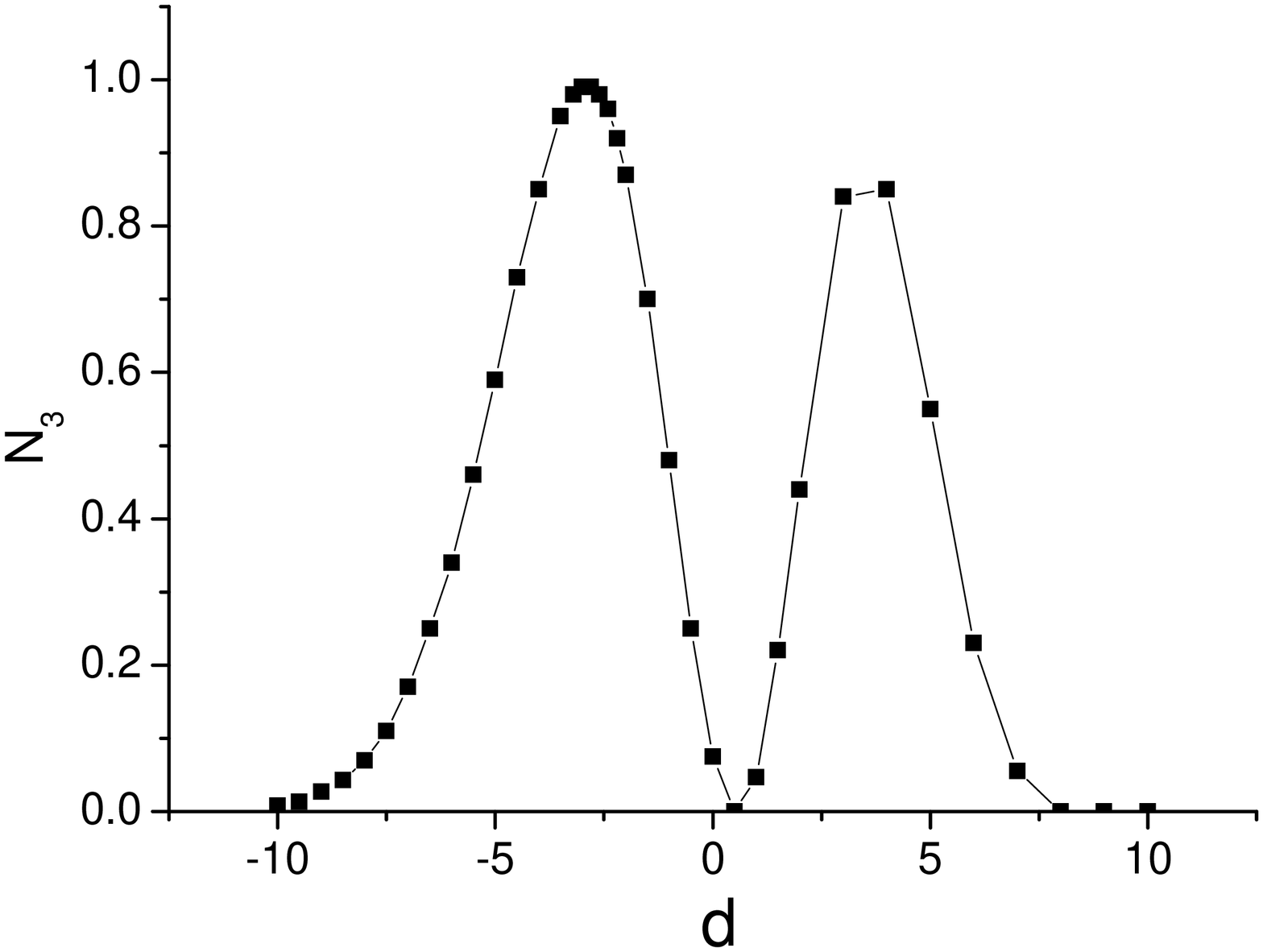}
\vspace{3mm}
\caption{\label{fig:shift}
Population $N_3$ at different relative shift $d$ between
pump and Stokes couplings. The maximal populations at counterintuitive $(d<0)$
and intuitive $(d>0)$ coupling order correspond to STIRAP and non-adiabatic
transport.
}
\end{figure}

The next panels c)-e) show that switching on the interaction
worsens the transport. The damage is negligible for $\Lambda=0.1$ and does not exceed
20$\%$ for $\Lambda=\pm 0.2$ but STIRAP ruins for larger interaction (not shown).
In any case, STIRAP is robust under a modest interaction
and this holds without any detuning. The comparison of the cases
d) and e) shows that the result somewhat depends on the interaction sign.

The remaining three panels f)-h) demonstrate role of detuning in the non-interacting
condensate ($\Lambda=0$). It is seen that a considerable detuning spoils the transport
and the result slightly depends on the detuning sign. Our calculations
generally confirm that a weak detuning is not harmful and, in accordance with
\cite{Graefe_PRA_06,Rab_PRA_08,Opat_arXiv_08}, may be even useful to amend slightly
adiabaticity of the process. However, the detuning is obviously not obligatory.
Moreover, in real conditions, the adiabaticity is never fully kept.
Nevertheless, the adiabatic transport schemes should work if diabatic
perturbations are not strong, which is confirmed by our results.

In Fig. 4 the dependence of the transport on the relative shift $d$ of the
pump and Stokes couplings is demonstrated.
The results are obtained for $\Lambda=\Delta=0$ and $\Gamma $=4.36.
Both counterintuitive (Stokes precedes pump, $d<0$) and intuitive (pump precedes Stokes,
$d>0$) sequences of the couplings are covered. It is seen that the best result
(complete transport) takes place for the counterintuitive order, $d = -3$, pertinent
for STIRAP. The intuitive order also leads to the appreciable
population, $N_3 \sim 80-90\%$ at $d=3$, though this transfer is not adiabatic.
What is remarkable, there is no any transfer without the shift, i.e. for $d=0$.
Being adiabatic, STIRAP transfer is less sensitive to the parameters of the process
and so is more preferable than its non-adiabatic counterpart. This is partly
confirmed by Fig. 4 where the left adiabatic peak is broader than the right
non-adiabatic one, hence less sensitivity to the shift $d$.

In Fig. 5 the phases $\phi_i$  and phase differences $\theta_i$ are given
for the cases with and without the interaction $U$. It is seen that the
interaction and corresponding non-linear effects drastically change
both $\phi_i$  and $\theta_i$. This conclusion generally agrees
with the observations for the oscillating BEC \cite{Smerzi_PRL_97,Raghavan_PRA_99}
where the interaction also strongly affects the phases. So the
interaction can in principle be implemented (via the Feshbach resonance) to
control geometric phases generated during BEC transport.

\subsection{Geometric phases}

%%%%%%%%%%%%
% Figure 5
%%%%%%%%%%%%
\begin{figure}
\includegraphics[height=9cm,width=6.0cm,angle=-90]{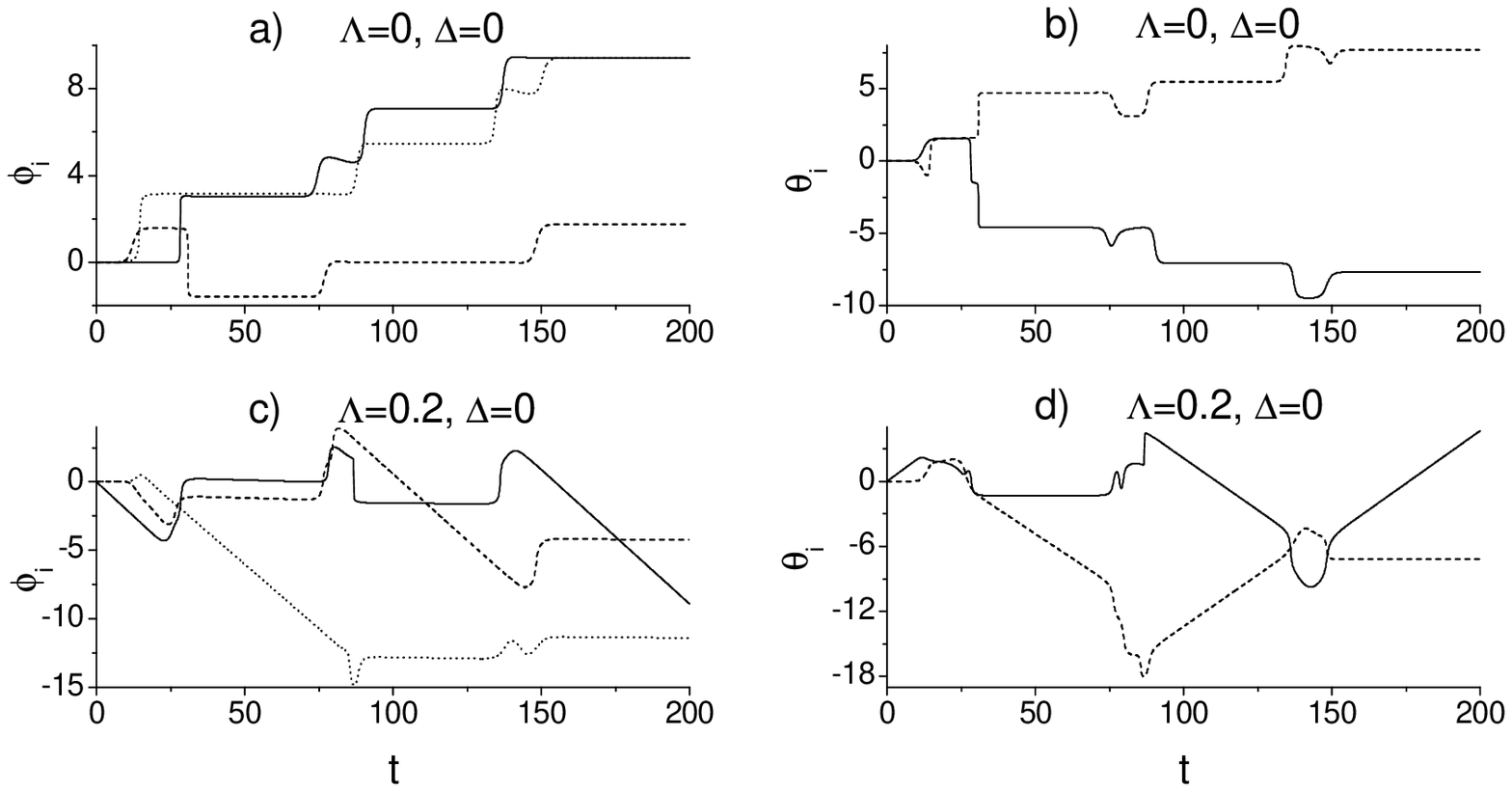}
\vspace{3mm}
\caption{\label{fig:phases}
Phases $\phi_i$ (i=1,2,3, left panels) and phase differences $\theta_i$
(i=1,2, right panels) calculated without ($\Lambda =0$, upper panels) and
with ($\Lambda =0.2$, bottom panels) interaction.
In all the panels $\Delta=0$. For i=1,2,3 the solid, dash, and dotted
lines are used. The phases in the upper and bottom panels correspond to the
populations in Figs. 3b) and 3d), respectively.
}
\end{figure}
%%%%%%%%%%%%
% Figure 6
%%%%%%%%%%%%
\begin{figure}
\includegraphics[height=9cm,width=4.1cm,angle=-90]{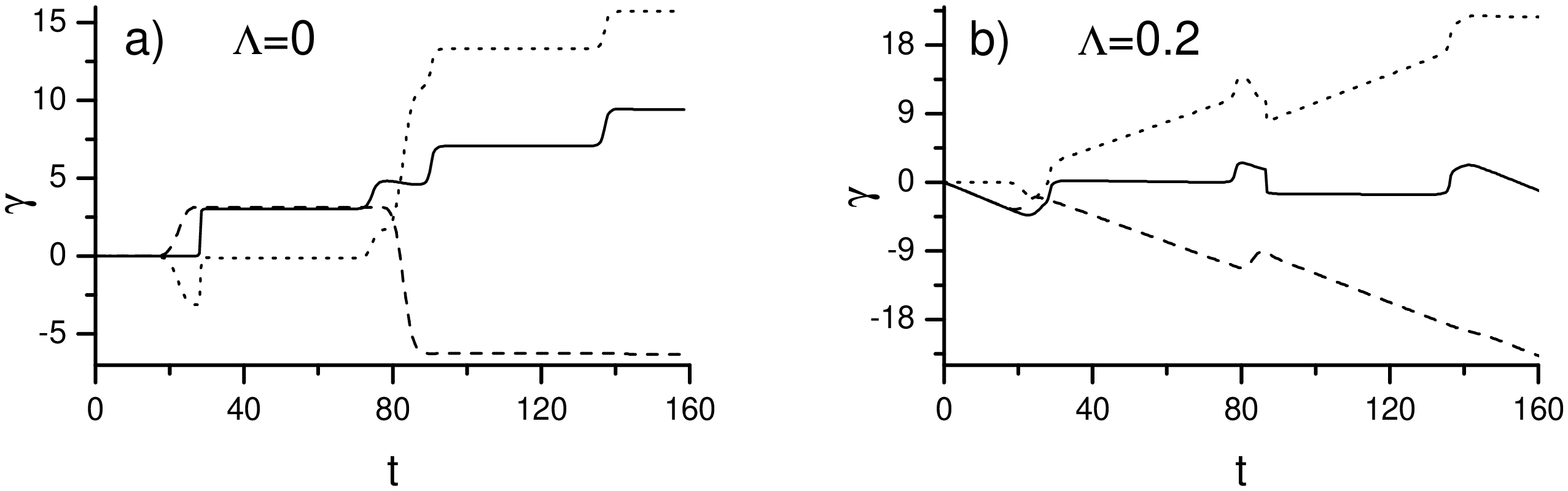}
\vspace{4mm}
\caption{\label{fig:geom}
Total $\gamma_t$  (solid line), dynamical $\gamma_d$ (dash line),
and geometric $\gamma_g$ (dotted line) phases
%in three-step STIRAP
for the cases without (a) and with (b) interaction $U$. In both panels
$\Delta=0$. In the panel a) the unconventional geometric phase
$\gamma_g=\alpha\gamma_d$ with $\alpha \ne -1$ is obtained.}
\end{figure}

Being coherent, BEC provides an interesting possibility to generate
and investigate various geometric (topological) phases $\gamma_g$
\cite{Fuentes_PRA_02}-\cite{Balakrishnan_EPJD_06}. These phases are
known to be mainly determined by the path topology (unlike the
dynamical phases $\gamma_p$ which depend on the process rate as well),
see e.g. \cite{Mukunda}. Because of this feature, $\gamma_g$ is less
sensitive to parameters of the process and so can serve as a reliable
carrier of information \cite{Nayak_RMP_08}. During last decades
implementation of geometric phases in the so called geometric quantum
computation has become actual \cite{Nayak_RMP_08_gp,Zhu_PRL_03_gp,Feng_PRA_07_gp}.
In this connection using BEC transport for exploration of various $\gamma_g$
could be indeed of a keen interest.

The geometric phases in BEC dynamics have been already studied elsewhere
\cite{Fuentes_PRA_02}-\cite{Balakrishnan_EPJD_06}. However, these studies
concerned the oscillating BEC and, by our knowledge,  still nothing was
done for $\gamma_g$ in BEC transport problems.

In principle, geometrical and dynamical phases in the oscillation and
transport dynamics can be described within the same general formalism
given in \cite{Mukunda}. In the present study, we exploit its version
\cite{Balakrishnan_EPJD_06}. There, for the cyclic evolution during
the time interval $t=[0,T]$, the geometric phase is determined as the
difference between  the total and the dynamical phases
\begin{equation}
 \gamma_g=\gamma_t-\gamma_d
\end{equation}
where
\begin{eqnarray}
  \gamma_t &=& \mbox{arg} \;(\vec\psi(0) \cdot \vec\psi(T)) \; ,
\\
  \gamma_d &=& \mbox{Im} \int^T_0 dt (\vec\psi(t) \cdot \dot{\vec\psi}(t)) \; ,
\end{eqnarray}
and
\begin{equation}
\vec\psi(t)=(\psi_1(t),\psi_2(t),\psi_3(t))
\end{equation}
is the state vector consisting of the components (\ref{order_par})
of the condensate.

In Fig. 6 results of our calculations of $\gamma_t$, $\gamma_d$, and $\gamma_g$
for the three-step STIRAP transport are presented. Due to a
cyclic adiabatic evolution, the Berry phase \cite{Berry} is produced.
The cases with and without interaction (corresponding to the
protocols of Figs. 3b), 3d) and 5) are considered.
It is seen that in both cases we observe proportionality of the geometric
and dynamical phases, $\gamma_g \approx \alpha \gamma_d$. However,
for $\Lambda=0.2$ there is a large mutual compensation of the phases
($\alpha \approx -1$) thus giving $\gamma_t \approx 0$. Instead for $\Lambda=0$
$\alpha \approx -2$ (for $t > 90$) and we gain the so called unconventional
geometric phase \cite{Zhu_PRL_03_gp}. This gives a chance to determine $\gamma_g$
in interference experiments via measurement of $\gamma_t$.
Indeed, if $\gamma_t \ne 0$ and we know $\alpha$ (which hopefully slightly depends
on the parameters of the process), then we directly get $\gamma_g$.

It worth noting that $\gamma_g \ne 0$ only if the path has a non-trivial topology.
Such topology becomes vivid if we reformulate the theory in terms of the $su(3)$
generators, in analogy to the $su(2)$ quasi-spin operators  treated e.g. in
\cite{Fuentes_PRA_02}.  In that two-mode case the  BEC dynamics
is reduced to the motion of the tip of the quasi-spin vector on the surface
of the Bloch sphere. Hence the non-trivial topology of the path (e.g. as
compared with motion on the plane). In such presentation, $\gamma_g$ is
determined by the solid angle subtended by the closed curve on the sphere surface.
The similar representation can be built for BEC transport in the triple-well
trap as well.

Altogether, BEC transport within STIRAP protocol seems to be a useful tool to
generate different geometric phase. Such transport can be realized for a variety
of the process parameters \cite{Rab_PRA_08,Opat_arXiv_08}. So, a
manifold of geometric phases can be produced. In this connection, it would be
interesting to look for the STIRAP protocol leading to $\gamma_t \ne 0$ at
$\gamma_d=0$, i.e. for the conditions where only $\gamma_g$ is produced.

\section{Conclusions}

The Stimulated Raman Adiabatic Passage (STIRAP) is applied to irreversible
transport of the Bose-Einstein condensate (BEC) in the triple-well trap. The basic
features of STIRAP are sketched and analogy between two-photon and tunneling
STIRAP scenarios is discussed. The relevant formalism is presented and specified
for the transport problem.

The calculations are performed for the cyclic transport of BEC by using three
successive STIRAP steps. It is shown that STIRAP indeed produces a robust
and complete transport. Besides, it remains effective at modest interaction
between BEC atoms and related non-linearity of the problem. As compared with
the previous STIRAP studies \cite{Graefe_PRA_06,Rab_PRA_08,Opat_arXiv_08}, we
demonstrate that detuning (trap asymmetry)  is not obligatory and, at its large
magnitude, can be even detrimental (though small detuning can slightly amend
adiabaticity of the process).

Note that full adiabaticity of STIRAP can be hardly ensured in BEC transport
since the transferred atoms must in any case pass the intermediate well thus
disturbing the adiabatic following. In this connection, we do not pursue the
perfect adiabaticity. Instead, we demonstrate that complete and robust
transport can be realized even under its (though modest) distortion.
Moreover, we show that effective transport can take place even at intuitive
sequence of partly overlapping couplings when the process is strictly
non-adiabatic. Note that at zero interactions our results are relevant for
the transport of individual atoms.

For the first time, we demonstrate evolution of phases of BEC fractions
in STIRAP transport and show that they strictly depend on the interaction.
The corresponding dynamical and geometric phases are also computed. It is
shown that at some interaction we gain the unconventional topological
phase which is proportional to its
dynamical counterpart and both them produce a large total phase. This finding
may be used to determine unconventional topological phases by measuring the total
phase in interference experiments. Altogether, our study show that STIRAP
transport can be a perspective tool for generation and exploration of various
geometric phases which in turn are now of a keen interest for quantum computing
\cite{Nayak_RMP_08_gp,Zhu_PRL_03_gp,Feng_PRA_07_gp}.

\begin{acknowledgments}
The work was partly supported by grants PVE 0067-11/2005 (CAPES, Brazil)
and 08-0200118 (RFBR, Russia). V.O.N. thanks Profs. V.I. Yukalov and
A.Yu. Cherny for useful discussions.
\end{acknowledgments}


\begin{thebibliography}{99}
\bibitem{note}
The basis is formed from the oscillator ground state
wave functions with shifted coordinates,
$\Phi_k ({\vec r})= u_({\vec r}-{\vec r}_k)$,
where ${\vec r}_k$ marks the minimum of the isolated k-th oscillator well.
For separated wells we have  $\int d\vec r \Phi^*_k\Phi_j = \delta_{kj}$.
\end{thebibliography}
\end{document}